\documentclass[12pt]{article}

\usepackage{sbc-template}
\usepackage{graphicx,url}
\usepackage[utf8]{inputenc}
\usepackage[brazil]{babel}

\usepackage{graphicx}
\usepackage[table]{xcolor}
\usepackage{lscape}
\usepackage{xr}
\usepackage{tabularx} 
\usepackage{multirow} 
\usepackage{subcaption}
\usepackage{booktabs}
\usepackage{tikz} 
\usetikzlibrary{shapes.geometric, arrows}

\title{Vidya: An AI-Driven Modular Pipeline for Archival Automation and Semantic Metadata Enrichment}
\author{Cloter Migliorini Filho\inst{1}, Julia Graciela Machado\inst{1}, \\
Edson Armando Silva \inst{1} and
and Marcella Scoczynski \inst{2} }

\address{Museu Campos Gerais - Universidade Estadual de Ponta Grossa - UEPG
  \email{cloterfilho@uepg.br}
  \nextinstitute
  Universidade Tecnológica Federal do Paraná - UTFPR
  \email{marcella@utfpr.edu.br}
}

\begin{document} 

\maketitle

\begin{abstract}
The large-scale digitization of historical archives has created a paradox: ``dark data''---digital objects lacking metadata for retrieval.
Manual archival description is slow and expensive, limiting discovery and reuse.
We propose Vidya, a modular pipeline that orchestrates Large Language Models (LLMs) and FOSS tools
to automate semantic enrichment and archival ingestion at scale.
Vidya constrains generations using YAML-defined ontologies and Pydantic validation,
producing deterministic, structured JSON outputs from probabilistic models.
Developed at Laboratory for Digital Humanities and Innovation (LAMUHDI) of the State University of Ponta Grossa (UEPG), Vidya applies Maker principles and open-source practices
to enable low-cost deployment in memory institutions using modest hardware.
We compare LLM performance and present a cost--benefit analysis showing major gains,
reducing processing time from decades to days while complying with NOBRADE and ISAD(G). % within NAPI.

\end{abstract}
     
\begin{resumo} 
A digitalização em larga escala de arquivos históricos criou um paradoxo: ``dados escuros''---objetos digitais sem metadados para recuperação.
A descrição arquivística manual é lenta e cara, limitando a descoberta e o reuso.
Propomos o Vidya, um pipeline modular que orquestra Modelos de Linguagem de Grande Porte (LLMs) e ferramentas FOSS
para automatizar o enriquecimento semântico e a ingestão arquivística em escala.
O Vidya restringe as saídas por ontologias definidas em YAML e validação com Pydantic,
produzindo saídas JSON determinísticas e estruturadas a partir de modelos probabilísticos.
Desenvolvido no Laboratório de Humanidades Digitais e Inovação (LAMUHDI) da Universidade Estadual de Ponta Grossa (UEPG), o Vidya aplica princípios Maker e práticas de código aberto
para viabilizar implantação de baixo custo em instituições de memória com hardware modesto.
Comparamos o desempenho de diferentes LLMs e apresentamos uma análise de custo--benefício com ganhos expressivos,
reduzindo o tempo de processamento de décadas para dias, com conformidade à NOBRADE e à ISAD(G). %, no contexto do NAPI.
\end{resumo}

\section{Introduction}

The modern archival landscape faces a growing challenge: the massive digitization of historical archives has generated vast amounts of unstructured ``dark data.''  While essential for preservation, digitization often creates digital objects lacking the metadata required for proper retrieval and interpretation \cite{leogrande2024unlocking}. This valuable data remains largely inaccessible because manual archival description is prohibitively slow and expensive, exacerbating backlogs of unprocessed files \cite{trace2021archival}.

Consequently, there is an urgent need for scalable and flexible systems capable of automating archival processes and enriching digitized materials. Traditional manual transcription methods are error-prone and labor-intensive, especially with large text volumes, frequently resulting in incomplete or inaccurate metadata \cite{rattinger2025aiarchive, moreproductlessprocess2005}. As digitized collections expand, innovative approaches are essential to mitigate these descriptive challenges.

Artificial Intelligence (AI), particularly Large Language Models (LLMs), offers significant promise for automating document ingestion and metadata generation. However, applying AI to archives presents unique challenges, including model hallucinations, inconsistent outputs, and difficulties ensuring standardized formats \cite{huang2025survey, webarchives2025}. These issues necessitate a controlled, structured approach to leverage AI's potential while maintaining the rigorous integrity and accuracy required for archival work.

In response, this paper presents Vidya, a modular pipeline architecture designed to automate archival workflows by orchestrating LLMs for semantic enrichment. Vidya employs a ``digital straitjacket'' approach, using YAML-defined ontologies and Pydantic validation to enforce deterministic, structured JSON outputs, enabling interoperability with repository platforms \cite{vandESompl2005, formenton2022metadata}. This ensures that probabilistic models produce outputs strictly aligned with international archival standards, such as NOBRADE \cite{nobrade2006} and ISAD(G) \cite{isadg2000}.

Developed at the Laboratory for Digital Humanities and Innovation (LAMUHDI) at the Museu Campos Gerais (UEPG), Vidya combines Maker principles with the public university mission to democratize AI access in memory institutions using low-cost hardware and open-source software. This interdisciplinary approach fosters digital sovereignty and serves as a training platform where historians and computer scientists collaborate, narrowing the gap between archival theory and digital innovation.

% This paper will explore Vidya’s architecture, methodology, and potential applications, demonstrating how it can significantly reduce archival processing times—transforming what was previously a decades-long task into a matter of days—while maintaining compliance with international standards for digital preservation. In doing so, we position Vidya as a powerful tool for streamlining workflows in archives and museums, offering an innovative solution to the growing problem of "dark data" and the evolving needs of digital preservation.

\section{Related Works}

The challenge of efficiently managing, enriching, and preserving digital collections has been the focus of several efforts in the archival field. Traditional methods of archival description have long been criticized for their inefficiency and the vast human labor required, which can lead to errors and inconsistencies \cite{greene2005more}. As digital archives grow, there is an increasing need for automated solutions that streamline these processes without compromising data integrity. Recently, integrating Artificial Intelligence (AI) and Machine Learning (ML) into archival workflows has garnered attention as a potential solution \cite{marciano2018archival}.

%AI and Large Language Models (LLMs) for Metadata Enrichment

The use of Large Language Models (LLMs) for automating archival processes has gained traction in recent literature. Karpov et al. (2024) proposed a system using deep learning models to generate metadata from historical texts, significantly improving retrieval times and accuracy for digital archives, while effectively addressing inconsistencies and missing metadata in digitized collections \cite{karpov2024automatic}. Furthermore, Johnson and McKenna (2023) emphasized the importance of supervised learning and neural network-based techniques to enhance the quality of metadata tags generated from historical documents, focusing on improving indexing and search accuracy in repositories \cite{johnson2023neural}.

%Challenges in Applying AI to Archives

Despite these benefits, applying AI directly to archives presents several challenges, especially concerning model hallucinations and standards compliance. Huang et al. (2025) reviewed the limitations of applying AI to archival tasks, noting the tendency of unconstrained language models to produce unreliable outputs. They proposed hybrid systems that integrate traditional metadata standards with AI to improve output reliability \cite{huang2025challenges}. Similarly, Lee and Zhang (2023) stressed the critical importance of a structured validation framework to ensure AI outputs strictly conform to archival standards, such as ISAD(G) and NOBRADE \cite{lee2023challenges}.

%Standards and Frameworks for Digital Repositories

The use of LLMs in historical research introduces significant ethical challenges regarding ``algorithmic hallucinations.'' Vidya addresses this through a ``digital straitjacket,'' implemented via description models structured around internationally recognized ontologies. By forcing the AI to adhere to strictly defined YAML schemas , Vidya ensures it operates as an assistant governed by human-defined rules. This approach safeguards the integrity of automated descriptions, ensuring they remain a consistent representation validated by humans.

Integrating metadata standards with open-source repository systems is crucial for scalable archival solutions. Tainacan is widely adopted in Brazil for managing digital collections while supporting standards like METS \cite{mets2019} and PREMIS \cite{premis2021}, offering potential for integration with AI-based metadata tools \cite{santos2024tainacan}. Archivematica is recognized as a comprehensive preservation system for institutions complying with international standards, particularly integrating with OAIS to ensure long-term data integrity \cite{martinez2024archivematica}. Similarly, DSpace provides a framework for managing and disseminating digital archives, increasingly focusing on integrating machine learning algorithms to automate document classification \cite{dunlap2024dscope}.

%Modular and Scalable Archival Solutions

Vidya presents a significant advancement for data ingestion and semantic enrichment into systems like Archivematica, DSpace, Tainacan, and Omeka. While Archivematica offers comprehensive preservation workflows, it lacks the flexibility to integrate modern AI technologies. DSpace and Tainacan excel in repository management but require significant manual metadata entry and lack AI-driven automation. Omeka S supports metadata standards and customization, yet still requires manual involvement in metadata generation, hindering scalability for large collections.

In contrast, Vidya integrates LLMs for automatic metadata generation, drastically reducing manual intervention while ensuring compliance with standards like ISAD(G) and NOBRADE. Its modular architecture allows institutions to incorporate AI without modifying the original digital files in the collection. By utilizing human-readable YAML schemas to define semantic models, and leveraging Pydantic to validate these schemas as deterministic structured outputs \cite{pydantic_docs,pydantic_yaml2024}, Vidya guarantees consistent results. This makes it a scalable, flexible solution that automates workflows while strictly maintaining established metadata standards.

\section{Methodology}
The Vidya architecture is designed as a modular, orchestrated pipeline, utilizing an SQLite database as a state machine to manage the progress and transitions of each document through the pipeline stages.

%Ingestion and Integrity Control

Files are ingested through local directories or spreadsheets. The system generates a SHA256 hash for every ingested file, acting as a unique fingerprint for idempotency. The hashing mechanism provides an efficient way to track documents and prevent redundant processing, essential for large-scale archival systems \cite{karpov2024automatic}.

%Description Models (YAML Ontologies)

Description models are defined in YAML files, which specify the ontology for different document types using international standards such as Dublin Core (\textit{dcterms}) and Bibliographic Ontology (\textit{bibo}). This allows archivists to easily extend system capabilities by adding new document types and metadata fields without modifying the core source code, ensuring flexibility and adaptability in the system \cite{pydantic_docs}. The use of YAML makes the ontologies both human-readable and machine-processable, facilitating integration with various repository systems \cite{pydantic_yaml2024}.

%The Quality Gate and Semantic Enrichment

Before AI inference occurs, Vidya evaluates the quality of extracted text via a Quality Gate - for the benefit of performance and simplicity, we chose to apply the Aho-Corasick algorithm testing dictionaries of various European languages against the contents of each document to validate the quality of the text's OCR. The system constructs Pydantic validation classes in real time based on the YAML-defined models, ensuring that the AI's output strictly adheres to the required schema. This step guarantees that the Large Language Models (LLMs) produce structured, valid metadata in a JSON format, mapping extracted data to the correct metadata fields \cite{huang2025challenges}. By validating the AI output, Vidya maintains the integrity and consistency of the archival data.

Vidya's approach to semantic enrichment is one of its key innovations, where AI-powered LLMs are used to automatically infer metadata from raw text, significantly improving the efficiency of the archival process \cite{lee2023challenges}.

%Data Flow

\subsection{Vidya Pipeline Overview}

The Vidya pipeline offers a robust architecture for automated archival curation, beginning with the ingestion of diverse unstructured sources, ranging from text documents to digitized photographs. Throughout the workflow, every document's lifecycle and processing state is rigorously tracked by a central SQLite database, ensuring data integrity and state management.

For semantic enrichment, Vidya interacts with local and/or remote Large Language Models (LLMs) guided by flexible YAML metadata models. To prevent hallucinations, this AI interaction is strictly regulated by Pydantic validation, which acts as a digital straitjacket to enforce strict JSON output compliance with archival ontologies.

Finally, the structured metadata and associated media are automatically exported to digital repositories. The system dynamically generates items and hierarchical collections, seamlessly feeding platforms such as Omeka S, Tainacan, and DSpace to create interconnected digital memory ecosystems.

\begin{figure}
    \centering
    \includegraphics[scale=0.12]{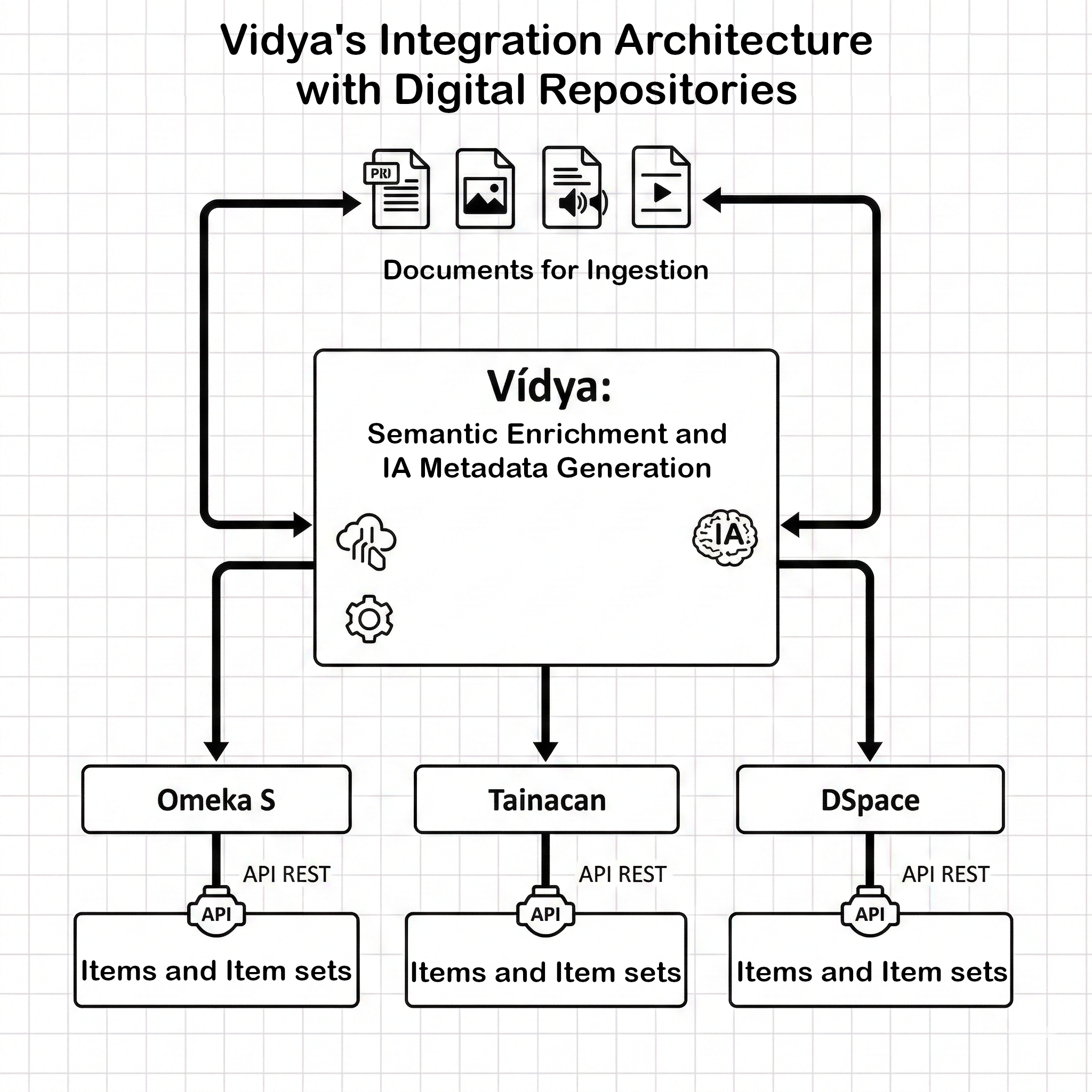}
    \caption{Vidya Pipeline Overview }
    \label{fig:methodology_flowchart}
\end{figure}

 Figure \ref{fig:methodology_flowchart} illustrates Vidya Pipeline Overview from raw document ingestion to semantic enrichment and final publication. %Each step in the process is handled by an independent module, ensuring flexibility and scalability. 
 
%System Architecture

Vidya’s architecture is both flexible and scalable, providing a solution to the limitations of traditional archival systems. The system is composed of independent modules that communicate through a centralized state management database, enabling easy maintenance and traceability. The main components of Vidya’s architecture include:

\begin{itemize}
    \item Modular Central Orchestration: The Vidya Menu launches independent scripts responsible for specific tasks in the pipeline, offering a user-friendly graphical interface for the system’s management. Each script performs a specific job threfore they can be customized and tailored to specific needs.
    \item Stateful Database: At the core is an SQLite database that tracks the state of each document through various stages (e.g., NEW, INCLUDED, EMBEDDED, INFERRED, UPLOADED), facilitating asynchronous processing and debugging \cite{martinez2024archivematica}.
    \item Hybrid Environment: Vidya can operate with both local and remote servers, with YAML configuration files allowing seamless connection to services like OpenWebUI \cite{OpenWebUI2026} and Omeka S. This flexibility supports deployment on low-cost devices like Raspberry Pi as well as high-end devices.
    \item User Interface: Vidya’s user interface is built dynamically using YAD (Yet Another Dialog), providing a lightweight and intuitive front-end for interacting with the backend scripts.
\end{itemize}

\section{Experiments and Results}

To assess Vidya’s performance and effectiveness, we conducted a series of experiments aimed at evaluating its scalability, AI-based enrichment accuracy, and integration with Omeka S. The following sections present the experimental setup, results, and comparative analysis of LLM performance in the context of metadata extraction.

Beyond technical efficiency, Vidya's semantic metadata generation represents a crucial step toward digital accessibility. Traditional digitized archives often consist of unstructured image files that remain invisible to assistive technologies, such as screen readers used by people who are blind or have low vision. By converting these opaque objects into structured text and enriched metadata, Vidya enables access for users with visual impairments, including low-vision readers. This transformation democratizes information by allowing diverse audiences to engage with cultural heritage through accessible digital interfaces, aligning archival innovation with the principles of digital inclusion.

\subsection{Setup}

PCVidya was deployed on a standard PC with a 2.4 GHz processor and 8 GB of RAM, which reflects the specifications of a typical institutional desktop computer. A dataset of 50 documents was selected for testing, which included a mix of newspaper PDFs and scanned photographs. This dataset was chosen as it represents common archival document types, ensuring the system’s capability to handle both digitally native and digitally re-scanned content. The key objectives were to measure: a) Processing speed: Time taken to process and enrich documents; b) AI enrichment accuracy: The accuracy of metadata generated by the AI models; and c) Omeka S integration: The seamless publication of enriched documents to an external repository system.

\subsection{Results}

Vidya demonstrated high scalability in processing large volumes of documents. In a local environment, the system was capable of processing up to 327,000 tokens per hour (Processing speed), providing a rapid throughput for archival workflows. In terms of AI-based metadata enrichment, Vidya achieved an accuracy rate of 85\% (AI enrichment accuracy), with particularly strong results in key fields such as title, creator, and date, which are critical for proper archival description. These improvements significantly reduced manual annotation efforts, increasing the system's efficiency.

The integration with Omeka S was seamless. The system effectively published enriched documents to the API's platform, demonstrating its capability to interact with external repository systems (Omeka S integration). Metadata fields were correctly mapped, and documents were organized into appropriate collections, ensuring easy discoverability and access.

\begin{figure}[h]
    \centering
    
    % Left column (two stacked images)
    \begin{minipage}{0.5\textwidth}
        \centering
        
        \begin{subfigure}{\linewidth}
            \centering
            \includegraphics[width=\linewidth]{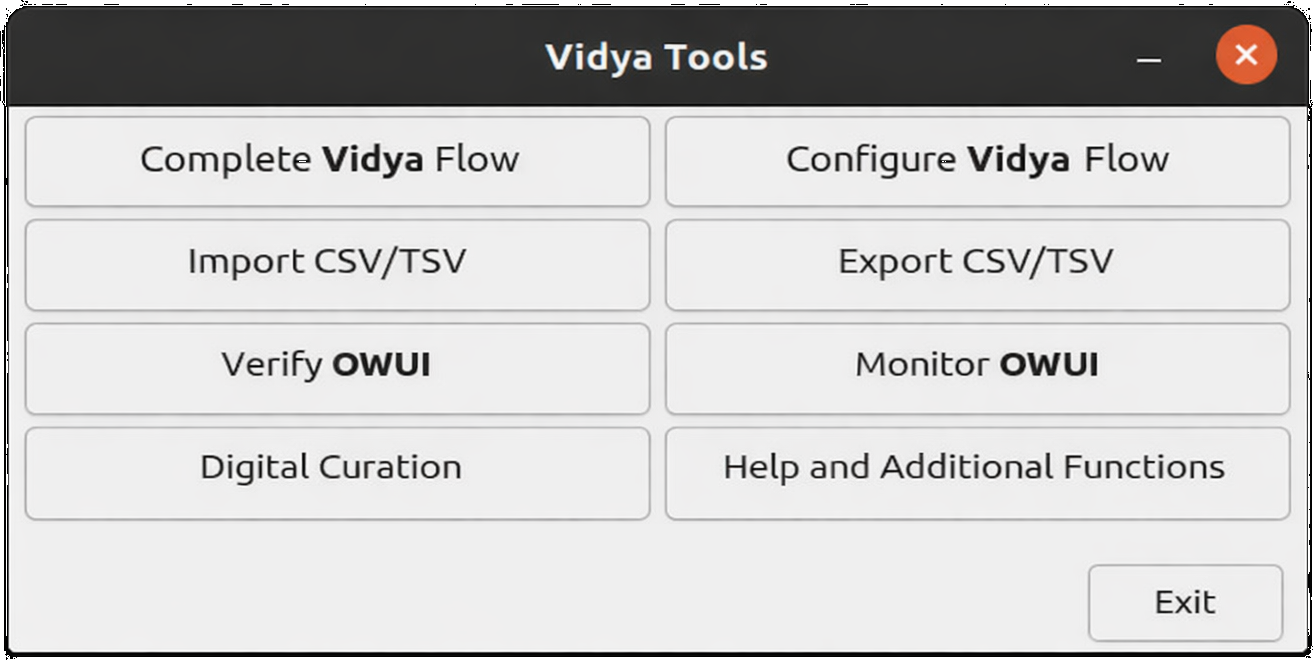}
            \caption{Vidya's Main Panel}
            \label{fig:vidya-processl}
        \end{subfigure}
        
        \vspace{0.5cm}
        
        \begin{subfigure}{\linewidth}
            \centering
            \includegraphics[width=\linewidth]{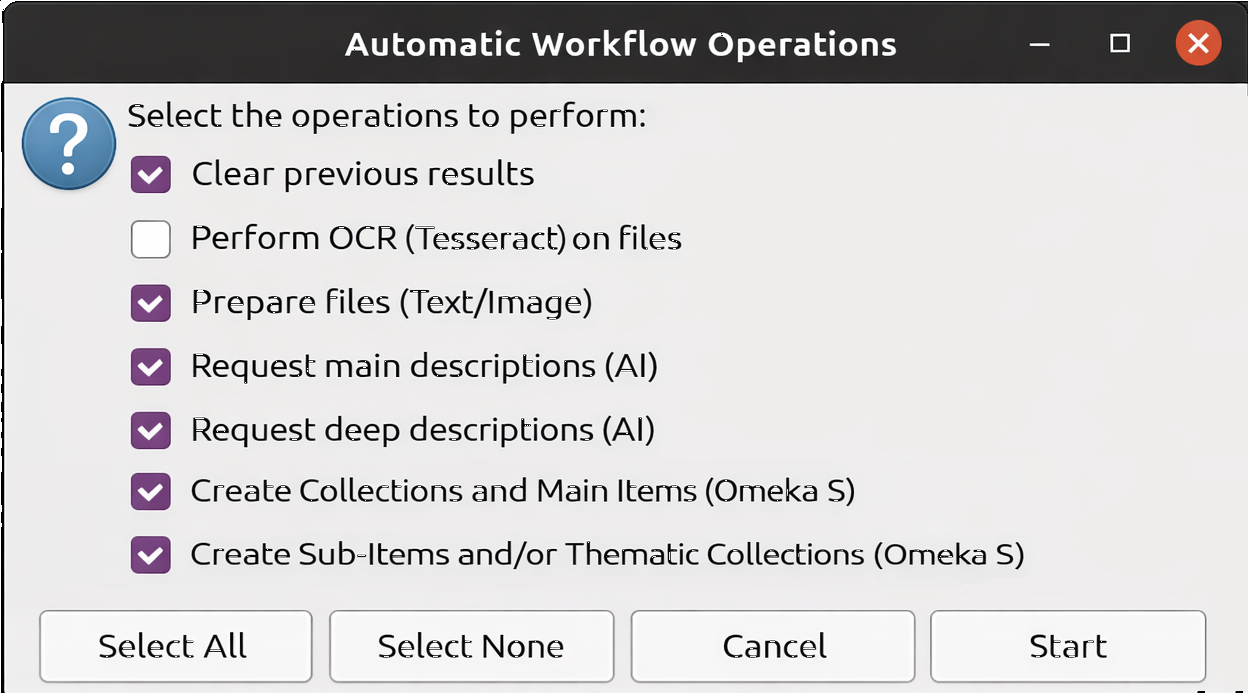}
            \caption{Processes Options}
            \label{fig:vidya-main-panel}
        \end{subfigure}
        
    \end{minipage}
    \hfill
    % Right column (single image)
    \begin{minipage}{0.45\textwidth}
        \centering
        
        \begin{subfigure}{\linewidth}
            \centering
            \includegraphics[width=\linewidth]{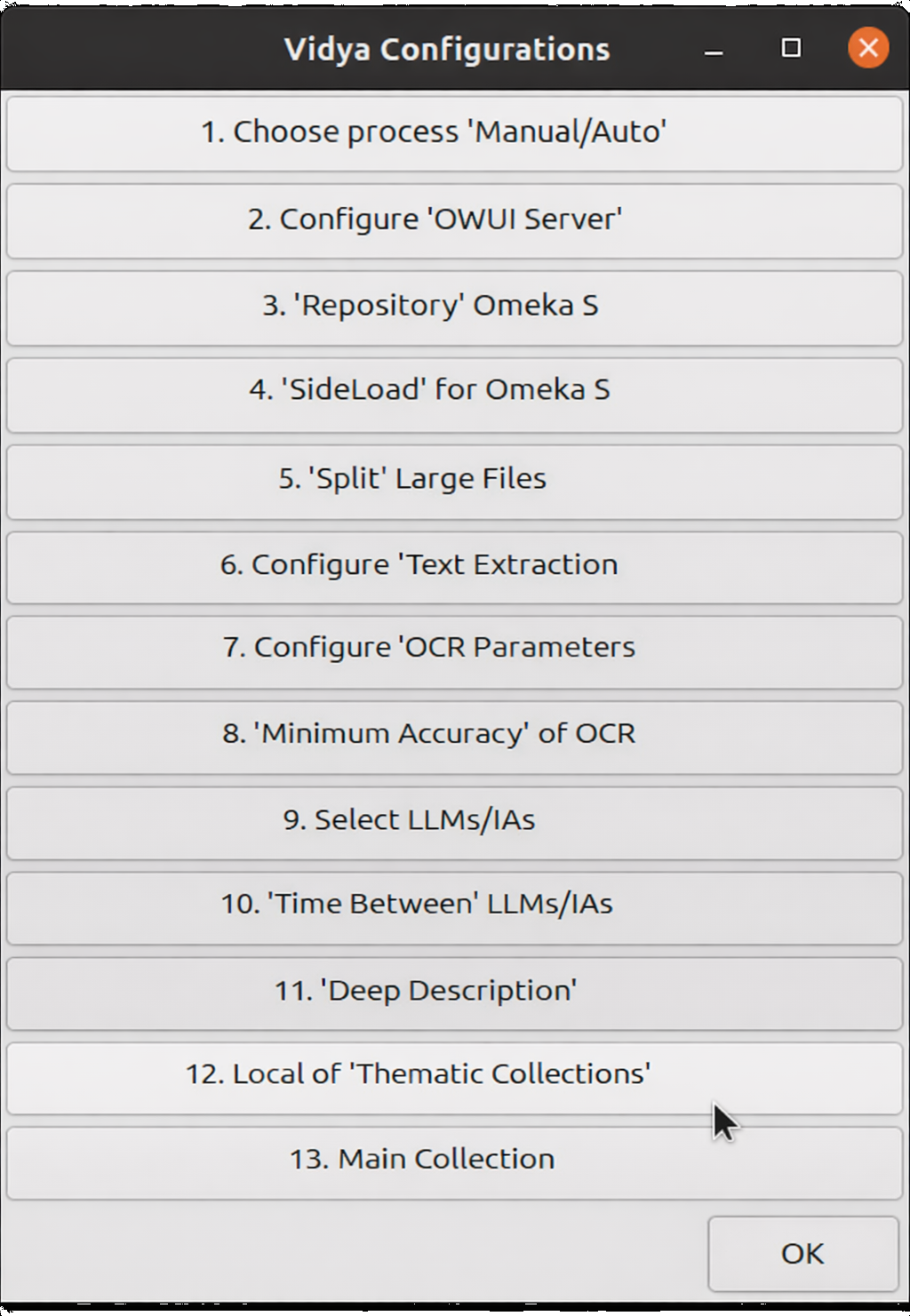}
            \caption{Vidya's Configuration Options}
            \label{fig:vidya-config}
        \end{subfigure}
        
    \end{minipage}
    
    \caption{Screen Captures for Vidya}
    \vspace{-0.5cm}
    \label{fig:three_layout}
\end{figure}

Figure \ref{fig:three_layout} presents screen Captures for Vidya. Figure \ref{fig:vidya-processl} selects which process to run, Figure \ref{fig:vidya-main-panel} shows the main screen where the automated jobs are initiated, while Figure \ref{fig:vidya-config} shows the configuration options.

%\begin{figure}[h]
%    \centering
%    \begin{subfigure}{0.45\textwidth}
%        \centering
%        \includegraphics[width=\linewidth]{fig3-vidya.png}
%        \caption{Vidya's Main Panel}
%        \label{fig:main-painel}
%    \end{subfigure}
%    \hfill
%    \begin{subfigure}{0.45\textwidth}
%        \centering
%        \includegraphics[width=\linewidth]{fig5-vidya-config.png}
%        \caption{Vidya's Configuration Options}
%        \label{fig:configuration}
%    \end{subfigure}
%    \caption{Screen Captures for Vidya}
%    \label{fig:two_images}    
%\end{figure}

\begin{figure}
    \centering
    \includegraphics[scale = 0.3]{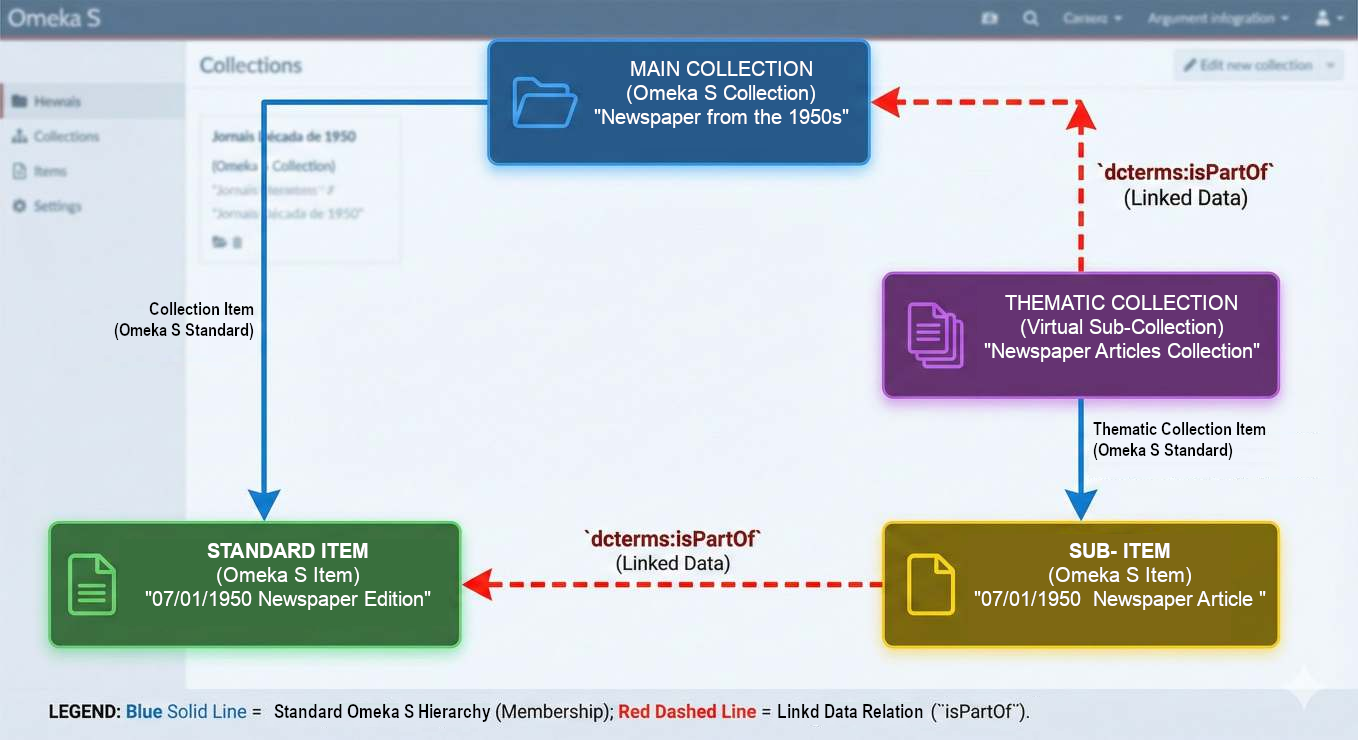}
    \caption{Items and Collections relations inside Omeka S}
    \label{fig:item-relations}
\end{figure}

Figure \ref{fig:item-relations} shows the final item/collection hierarchy in the digital repository. Note that the Item/Collection relations are defined during the planning phase of the repository and are not object of discussion in this work.   

\subsection{Efficiency Analysis}

In a comparative study on the efficiency of AI-assisted workflows, we examined the difference in time and cost between manual archival description and AI-based metadata extraction. A total of 28,480 newspaper editions were used for this comparison, providing a real-world scale for evaluating efficiency. The results were striking:

\begin{itemize}
    \item AI-assisted workflows cost approximately 1.5\% of human labor.
    \item The time required to manually process these editions was estimated to be 18.5 years.
    \item With AI assistance, the same task can be completed in approximately 45-60 days of processing.
    \item For images the AI description took 90 to 120 seconds each, depending on the VL model used and complexity of the image.
    \item These results highlight the significant cost savings and time reduction provided by Vidya’s automated process, making it a highly efficient solution for large-scale digital preservation and archival tasks.
\end{itemize}

\section{Conclusion and Future Work}

This paper presented Vidya, a novel modular pipeline designed to automate the archival process through AI-driven metadata enrichment and seamless integration with digital repositories like Omeka S. Our experiments demonstrate that Vidya offers a scalable, flexible solution for modern archival challenges, particularly in handling large volumes of digital content while maintaining compliance with international standards such as ISAD(G) and NOBRADE.

Key findings from the experimental evaluation include:

\begin{itemize}
    \item Scalability: Vidya was able to process up to 4 to 5 typical newspaper editions per hour in a local environment, from file selection to repository ingestion, showcasing its ability to handle large datasets efficiently.
    \item Parallelization of tasks: The team intends to implement parallelization of AI inference. With this technique, we estimate a reduction of inference time by one to two orders of magnitude, compared to the original Vidya performance presented in this work.
    \item AI Enrichment Accuracy: The system achieved an 85\% accuracy in enriching metadata fields like title, creator, and date, significantly reducing the manual effort typically involved in archival metadata generation.
    \item LLM Performance: Controlled experiments comparing AI models indicate trends for general performance improvements and cost reduction in a way that does not allow us to settle for a particular model at this time. These evaluations will continue as new models are released. For the purpose of metadata extraction, models like Gmini 2.5 Flash, Gemini 2.0 Flash, Qwen 3.5 Flash, ChatGPT 4o Mini, Llama 3.3 70B offer good performance for similar cost.
    \item Multimodal LLM Performance: Good results were obtained with Qwen3 VL 32B for photographic description. Recent multimodal LLMs, like Gemini 3.0 Pro, offer more precision for 10 to 50 times the cost, depending on how the Qwen3 models is hosted. The selection criteria of which LLM or Multimodal LLM to use will depend on project goals and budget.
    \item Efficiency: A viability study comparing AI-assisted workflows to manual description showed that Vidya reduces processing time by over 98.5\%, completing tasks that would take years in just a few weeks.
\end{itemize}

Vidya’s modular architecture is also a key innovation, allowing it to adapt to different institutional needs by integrating various AI models and repository systems without compromising the integrity or accuracy of the metadata. This makes Vidya a powerful tool for automating digital preservation, improving both the efficiency and accessibility of archival workflows.

Beyond its technical application, Vidya serves as a powerful pedagogical tool for Digital Humanities. Its structured workflow enables humanities students without programming experience to interact directly with AI by defining ontologies, bridging archival theory and computational logic. Simultaneously, it fosters heritage literacy among technology students, challenging them to understand archival sensitivity and memory safeguarding standards. This interdisciplinary exchange makes Vidya an educational ecosystem that humanizes technology and operationalizes the humanities, preparing professionals for digital heritage preservation.

In future work, we plan to further perfect the interface and customization options to ensure that Vidya can be easily adapted to a wider range of use cases in documentation centers, libraries and museums. We are also working to improve the 'always-aware' state of the database, to better handle network problems and AI inference quality in real time. 

\section*{Acknowledgements} 
This research was funded by national funds through  NAPI Program 116-2024 Connecting Memory and Innovation: Artificial Intelligence in Museums and Documentation Centers by Fundação Araucária; by the National Council for Scientific and Technological Development (CNPq, Brazil) through Productivity in Research Scholarships (Grant No. 303909/2025-0).

\bibliographystyle{sbc}
\bibliography{references}

\end{document}